\documentclass{aa}
\usepackage{graphicx}
\usepackage{natbib}
\usepackage{scalerel}
\usepackage{multirow}   
\usepackage{mathrsfs}
\usepackage[table]{xcolor}
\usepackage{fontawesome}
\usepackage{pifont}
\usepackage{lipsum}  
\newcommand{\concept}{{\textsc{C}\textsc{o\scalerel*{\textbf{\textit{N}}}{n}cept}}}

\newcommand{\class}{{\tt CLASS}}
\newcommand{\camb}{{\tt CAMB}}
\newcommand{\rockstar}{{\tt ROCKSTAR}}
\newcommand{\consistent}{{\tt CONSISTENT}}

\newcommand{\de}{\text{d}}
\newcommand{\cmark}{\ding{51}}
\newcommand{\xmark}{\ding{55}}

\definecolor{valecol}{rgb}{0,0.5, 1.}

\usepackage{txfonts}
\usepackage[pdfencoding=auto,psdextra]{hyperref}
\hypersetup{
    colorlinks=true,
    linkcolor=blue,
    filecolor=magenta,
    urlcolor=blue,
    citecolor=blue
}
\usepackage[utf8]{inputenc}
\usepackage{orcidlink}

\begin{document}

   \title{DUCA: Dynamic Universe Cosmological Analysis}
   \subtitle{II. The impact of clustering dark energy on the halo mass function}
   \titlerunning{The impact of clustering dark energy on the halo mass function}
   \authorrunning{Castro et al.}

   \author{T.~Castro\orcidlink{0000-0002-6292-3228}\inst{1,2,3,4}\thanks{\email{tiago.batalha@inaf.it}},
   S.~Borgani\orcidlink{0000-0001-6151-6439}\inst{5,1,2,3,4},
   J.~Dakin\orcidlink{0000-0002-2915-0315}\inst{6},
   V.~Marra\orcidlink{0000-0002-7773-1579}\inst{7, 1, 3},
   R.~C.~Batista\orcidlink{0000-0002-7655-8719}\inst{8}, and
   L.~Salvati\inst{9,1}
   }

   \institute{
   $^{1}$ INAF -- Osservatorio Astronomico di Trieste, Via G. B. Tiepolo 11, 34143 Trieste, Italy\\
   $^{2}$ INFN -- Sezione di Trieste, Via Valerio 2, 34127 Trieste TS, Italy\\
   $^{3}$ IFPU -- Institute for Fundamental Physics of the Universe, via Beirut 2, 34151 Trieste, Italy\\
   $^{4}$ ICSC -- Via Magnanelli 2, Bologna, Italy\\
   $^{5}$ Dipartimento di Fisica -- Sezione di Astronomia, Universit\`a di Trieste, Via Tiepolo 11, Trieste, 34131, Italy\\
   $^{6}$ Department of Astrophysics -- University of Zurich, Winterthurerstrasse 190, 8057 Z\"urich, Switzerland\\
   $^{7}$ Departamento de Física, Universidade Federal do Espírito Santo, 29075-910, Vitória, ES, Brazil\\
   $^{8}$ Escola de Ciências e Tecnologia, Universidade Federal do Rio Grande do Norte, Campus Universitário Lagoa Nova, 59078-970, Natal, RN, Brazil\\
   $^{9}$ Université Paris-Saclay, CNRS, Institut d’Astrophysique Spatiale, 91405, Orsay, France
             }

\abstract
{
    Galaxy clusters are powerful probes of cosmology, and the halo mass function (HMF) serves as a fundamental tool for extracting cosmological information. Previous calibrations of the HMF in dynamical dark energy (DE) models either assumed a homogeneous DE component or a fixed sound speed of unity, which strongly suppresses DE perturbations.
    }{
    We extend the HMF calibration to clustering dark energy (CDE) models by allowing for a sound speed $(c_{\rm s})$ value different than unity. This generalization enables a broader description of the impact of DE perturbations on structure formation.
    }{
    Our approach builds upon the DUCA simulation suite that accounts for DE at the background and perturbative levels. We present an HMF calibration based on introducing an effective peak height while maintaining the multiplicity function as previously calibrated. The effective peak height is written as a function of the peak height computed using the matter power spectrum of the homogeneous DE case, but it is modulated by the amplitude of DE and matter perturbations on the non-homogeneous case at the turnaround. The model depends on one single parameter, which we calibrate using \(N\)-body simulations, following a Bayesian approach.
    }{
    The resulting HMF model achieves sub-percent accuracy over a wide range of \(c_{\rm s}\) values. Our analysis reveals that, although the overall impact of CDE on halo abundances remains modest (typically a few percent), the effects are more pronounced in non-phantom DE scenarios. Our model qualitatively agrees with predictions based on the spherical collapse model, but predicts a significantly lower impact for low $c_{\rm s}$.
    }{
    Our results underscore the need for more precise modeling of CDE's nonlinear regime. Numerical simulations and theoretical approaches must be advanced to capture the complex interplay between DE perturbations and matter fully.
    }


\maketitle

\section{Introduction}
\label{sec:intro}

Structure formation in the Universe proceeds hierarchically. Small-scale perturbations collapse and merge to form progressively larger structures over cosmic time. Galaxy clusters, being the most massive virialized objects, occupy a prominent position in this hierarchy and serve as powerful probes of cosmology~\citep[see reviews by][]{Allen:2011zs, KravtsovBorgani:2012}. 

The halo mass function (HMF), which provides the comoving number density of matter halos as a function of mass and redshift, is a fundamental theoretical tool for interpreting observations of galaxy clusters. Due to the strong non-linearity behind the galaxy cluster formation and evolution, analytical modeling of the HMF is limited to a qualitative description that quantitatively does not fulfill the requirements needed for precision cosmology~\citep[e.g.,][]{Press:1973iz, Bond:1990iw}. To circumvent this, the community started developing physically motivated semi-analytical models~\citep[e.g.,][]{Sheth:1999mn, Sheth:1999su, Tinker:2008ff, Despali:2015yla, Ondaro-Mallea:2021yfv, Euclid:2022dbc} that, when calibrated on simulations, provide a much more accurate prediction of the halo abundances. 

The most comprehensive simulations to calibrate the HMF models would be cosmological hydrodynamical simulations, as they simulate gravity and baryonic effects associated with galaxy formation. Baryons, although sub-dominant to dark matter, are known to impact the cluster abundance in a significant way~\citep[e.g.][]{Cui:2014aga,Velliscig:2014bza,Bocquet:2015pva,Castro:2017tbn,Castro:2020yes,10.1093/mnras/stad2419,Euclid:2023jih}. However, their cost limits the number of simulations that can be used for a robust exploratory model calibration. Therefore, the community took the strategy of calibrating the HMF model on gravity-only simulations and adding the baryonic impact as post-processing. We will follow this strategy, too, and focus on gravity-only simulations in the rest of the paper.

Recent and forthcoming observational campaigns, such as those conducted by the Vera C. Rubin Observatory's Legacy Survey of Space and Time~\citep[LSST,][]{LSSTScience:2009jmu},\footnote{\url{https://www.lsst.org}} the third generation of the South Pole Telescope~\citep[SPT-3G,][]{SPT-3G:2014dbx},\footnote{\url{https://astro.fnal.gov/science/cmbr/spt-3g/}} eROSITA~\citep{eROSITA:2020emt},\footnote{\url{https://www.mpe.mpg.de/eROSITA}} the Square Kilometre Array~\citep[SKA,][]{Maartens:2015mra},\footnote{\url{https://www.skatelescope.org}} the Dark Energy Spectroscopic Instrument~\citep[DESI,][]{DESI:2016fyo},\footnote{\url{https://www.desi.lbl.gov}} the Nancy Grace Roman Space Telescope~\citep{Spergel:2015sza},\footnote{\url{https://roman.gsfc.nasa.gov}} and Euclid~\citep{euclidoverview}\footnote{\url{https://www.euclid-ec.org}} will provide measurements with unprecedented statistical precision. To fully exploit these data, theoretical models must keep pace and achieve sub-percent accuracy so systematic uncertainties remain subdominant.

In the standard cosmological model, the Universe's accelerated expansion is driven by a cosmological constant $\Lambda$, which dominates the current energy density composition of the Universe, followed by cold dark matter (CDM) and baryons. Altogether, these components correspond to more than 99 percent of the current Universe's energetic composition, with massive neutrinos and relativistic species closing the missing gap. \citet{Euclid:2022dbc} presented a calibration of the HMF that reproduces with sub-percent accuracy the outcome of simulations assuming a standard $\Lambda$CDM cosmological model. A straightforward extension for the cosmological constant~\citep[see, e.g.,][for reviews]{Peebles:2002gy, Copeland:2006wr} is considering a smooth dark energy (DE) component with a variable equation of state (EOS) and constant sound speed $c_{\rm s}=1$. Within this phenomenological approach, DE remains effectively homogeneous on sub-horizon scales, and its perturbations do not affect structure formation significantly. \citet{DucaI} (hereafter Paper~I) extended the HMF model of~\citet{Euclid:2022dbc} to dynamical DE models described by the Chevallier--Polarski--Linder (CPL) parametrization~\citep{Chevallier:2000qy, Linder:2002et}. 

Clustering dark energy (CDE) becomes significant on sub‐horizon scales when the DE sound speed $c_{\rm s}$ drops below unity, permitting its perturbations to grow rather than remain suppressed.  This regime was first quantified by \citet{Weller:2003hw} and \citet{Bean:2003fb}, who analysed its imprints on the cosmic microwave background and the matter power spectrum.  Explicit field‐theory realizations --- such as k‐essence and tachyon models --- were formulated by \citet{Garriga:1999vw}, \citet{Armendariz-Picon:2000nqq}, and \citet{Sen:2002nu}, demonstrating that $c_{\rm s}$ can deviate substantially from unity.  To assess non‐linear effects, \citet{Mota:2004pa} applied a spherical collapse model to inhomogeneous DE, finding consequential shifts in the collapse threshold. These analytic approaches were refined by \citet{Abramo:2008ip} and \citet{Chang:2017vhs}.  More recently, cosmological simulation codes, such as \citet{Dakin:2019vnj}, \citet{Hassani:2019lmy}, and~\citet{Blot:2022mnt}, were developed.  In~\citet{Dakin:2019vnj}, a linear‐realization ansatz is adopted, capturing the leading impact of DE perturbations on dark matter dynamics while omitting the backreaction of non‐linear matter clustering on the DE field.  In~\citet{Hassani:2019lmy}, the authors employ an effective field theory~\citep[EFT; see,][]{Gubitosi:2012hu} framework to evolve DE perturbations and dark matter in a coupled non‐linear scheme. Lastly, in~\citet{Blot:2022mnt}, DE is modeled as an effective fluid, and the associated continuity and Euler equations are solved using a hydrodynamic scheme.  

The non‐linear modeling of CDE remains an open research frontier~\citep{Creminelli:2008wc,Frusciante:2017nfr,Cusin:2017wjg,Cusin:2017mzw}. For instance, \citet{Hassani:2021tdd,Hassani:2022xyb,Eckmann:2022wtd} reported potential instabilities in the EFT description.  Due to the unresolved issues with fully non-linear schemes, we adopt the linear realization approach proposed by \citet{Dakin:2019vnj} hereafter and refer to~\citet{Batista:2021uhb} for a deeper discussion on CDE models and their effects on structure formation.

In \citet{Batista:2017lwf}, the authors show that a null sound speed introduces an additional degree of freedom that alters the growth of density perturbations. In scenarios where $c_{\rm s}<1$, DE perturbations are less suppressed on small scales, modifying the gravitational instability that drives the collapse of dark matter halos. This change affects key quantities in the spherical collapse model -- such as the critical density threshold~\citep[e.g.,][]{Press:1973iz, Bond:1990iw} and the virial overdensity~\citep[e.g.,][]{Sheth:1999mn, Sheth:1999su} -- and ultimately alters the shape of the HMF. Therefore, accurate calibration of the HMF in CDE cosmologies must explicitly account for the impact of a variable sound speed on the non-linear evolution of density perturbations.

The impending arrival of high-precision data underscores the necessity for improved theoretical modeling. As noted by~\citet{Salvati:2020exw} and~\citet{Artis:2021tjj}, ongoing and forthcoming surveys of galaxy clusters will probe the large-scale structure of the Universe with an accuracy that demands theoretical predictions be controlled at the sub-percent level. Extending the HMF's calibration beyond and incorporating the effects of DE clustering with a variable sound speed is imperative to meet these requirements.

In this work, we present a new HMF model that explicitly includes the impact of a variable sound speed of DE. We achieve this by extending the DUCA $N$-body suite of simulations presented in Paper~I to include models with $c_{\rm s}<1$. Remarkably, our new HMF model reproduces the results from the new CDE simulations without reducing the accuracy achieved before.

This paper is organized as follows: Section~\ref{sec:theory} describes the theoretical framework for CDE with a variable sound speed. Section~\ref{sec:methodology} details the simulation setup employed to model the extended HMF model. Section~\ref{sec:results} presents the main results, including the modeling, comparisons with previous HMF calibrations, and tests of model robustness. We discuss the implications of our results in Section~\ref{sec:discussion}. Finally, in Section~\ref{sec:conclusions}, we draw our conclusions.

\section{Theoretical framework}
\label{sec:theory}

In this section, we present a short overview of the main concepts of the CDE model adopted. We present the background evolution in Sect.~\ref{sec:background}, the perturbations in Sect.~\ref{sec:perturbations}, and the HMF formalism in Sect.~\ref{sec:hmf}.

\subsection{Background evolution}
\label{sec:background}

We adopt the CPL phenomenological framework to characterize the evolution of the DE EOS, $w_{\rm DE}$, which links DE's pressure $p_{\rm DE}$ and energy density $\rho_{\rm DE}$
\begin{equation}
w_{\rm DE}(a) = \frac{p_{\rm DE}}{\rho_{\rm DE}}\,.
\label{eq:w_def}
\end{equation}
In this approach, $w_{\rm DE}$ is modeled as a linear function of the scale factor $a$
\begin{equation}
w_{\rm DE}(a) = w_0 + w_a (1 - a)\,,
\label{eq:w_of_a}
\end{equation}
where \(w_0\) represents the present-day value of the EOS, and \(w_a\) quantifies its rate of change with the scale factor.

In a spatially flat universe that contains matter, radiation, massive neutrinos, and DE, the Hubble parameter \(H(z)\) is given by
\begin{align}
\frac{H^2(z)}{H_0^2} &= \Omega_{\rm m,0}\,(1+z)^3 + \Omega_{\rm r,0}\,(1+z)^4 + \Omega_\nu(z)\,\frac{\rho_{\rm c}(z)}{\rho_{\rm c,0}} \nonumber\\
&\quad +\, \Omega_{\rm DE,0}\,(1+z)^{3(1+w_0+w_a)}\,\exp\!\left[-\frac{3w_a\,z}{1+z}\right]\,,
\label{eq:hubble_parameter}
\end{align}
where \(H_0\) denotes the present-day Hubble constant; \(\Omega_{\rm m,0}\), \(\Omega_{\rm r,0}\), and \(\Omega_{\rm DE,0}\) are the current density parameters for matter (both baryonic and CDM), radiation, and DE, respectively. Here, \(\Omega_\nu(z)\) is the neutrino density parameter at redshift \(z\), while \(\rho_{\rm c}(z)\) and \(\rho_{\rm c,0}\) are the critical densities at redshift \(z\) and at present, respectively. While massive neutrinos also significantly impact cosmic evolution by influencing both the expansion history and the growth of structure \citep{Lesgourgues:2006nd}, we will not consider them in the rest of the paper as their impact on the HMF has already been studied~\citep{Castorina:2014, Costanzi:2013bha,Vagnozzi:2018pwo} and validated~\citep[][Paper~I]{Euclid:2022dbc}.

A key aspect of DE models is their ability to account for the observed accelerated expansion of the Universe. The acceleration is quantified by the deceleration parameter, \(q_{\rm dec}\), defined as \citep{Weinberg:2008zzc, peebles2020large}
\begin{equation}
q_{\rm dec} = -\frac{\ddot{a}}{a\,H^2}\,,
\label{eq:deceleration_parameter}
\end{equation}
with the overdots indicating derivatives with respect to cosmic time. By invoking the Friedmann equations, \(q_{\rm dec}\) can be expressed in terms of the density parameters and their associated EOS. At the current epoch (\(z=0\)), the deceleration parameter reduces approximately to
\begin{equation}
q_{\rm dec,0} \approx \frac{1}{2}\left[2\,\Omega_{\rm m,0} + \Omega_{\rm DE,0}\,(1+3w_0)\right]\,.
\label{eq:q0}
\end{equation}
A negative \(q_{\rm dec}\) signifies that the Universe is accelerating, which requires that \(w_0 < -1/3\).

\subsection{Dark energy perturbations}
\label{sec:perturbations}

As the nature of the DE component that drives the accelerated expansion remains unknown, a compelling description of DE is achieved by parameterizing its perturbations.  Two widely used approaches that have been implemented in both \class~\citep{Blas:2011rf} and \camb~\citep{Lewis:1999bs} are the fluid formalism and the parametrized post-Friedmann (PPF) formalism~\citep{Hu:2008zd,Fang:2008sn,Baker:2012zs}. The fluid formalism evolves DE density and velocity perturbations directly from an assumed equation of state and sound speed, offering transparent physical interpretation. The PPF framework, by contrast, enforces a stable interpolation across regimes where the DE equation of state crosses the phantom divide ($w=-1$), thus avoiding the divergences that plague simple fluid treatments in that case.  In our simulations, we will adopt one of these two prescriptions, both of which are available within the \concept\ code and encompass models such as quintessence-like scenarios.  In what follows, we briefly review both formalisms.  

\subsubsection{Dark energy as a fluid}

In the fluid formalism, DE is described by an EOS, $w(a)$, and a constant rest-frame sound speed, $c_{\rm{s}}$. The EOS adopted in this paper is described in Sect.~\ref{sec:background}; however, the discussion below is independent of the particular form of $w(a)$.

In the synchronous gauge, the continuity and Euler equations for the DE fluid are written in Fourier space as~\citep[see, e.g.,][]{Ballesteros:2010ks}
\begin{align}
{\delta}_{\rm{DE}}^{\prime\rm{s}} = &- (1+w) \left( \theta_{\rm{DE}}^{\rm{s}} + \frac{\mathfrak{H}^\prime}{2} \right)
- 3 \left( c_{\rm{s}}^2 - w \right) \mathcal{H} \, \delta_{\rm{DE}}^{\rm{s}}\nonumber\\
&- 9 (1+w) \left( c_{\rm{s}}^2 - c_{\rm{a}}^2 \right) \mathcal{H}^2 \frac{\theta_{\rm{DE}}^{\rm{s}}}{k^2} \,, \label{eq:fluid-continuity} \\
{\theta}_{\rm{DE}}^{\prime\rm{s}} &= - (1-3 c_{\rm{s}}^2) \mathcal{H} \, \theta_{\rm{DE}}^{\rm{s}}
+ \frac{c_{\rm{s}}^2 k^2}{1+w} \, \delta_{\rm{DE}}^{\rm{s}}
- k^2 \sigma_{\rm{DE}} \,, \label{eq:fluid-euler}
\end{align}
where a prime superscript denotes a derivative with respect to conformal time $\tau$, $\delta_{\rm{DE}}^{\rm{s}}$ and $\theta_{\rm{DE}}^{\rm{s}}$ are the DE density contrast and velocity divergence, respectively, and $\mathcal{H}\equiv a^\prime/a$, $\mathfrak{H}$ is one of the synchronous gauge metric perturbations\footnote{In the literature it is commonly used $h$ for representing this metric perturbation, but we instead use the fraktur font not to confuse it with the Hubble parameter $h=H_0/(100\,{\rm km}\,s^{-1}\,{\rm Mpc}^{-1}$).}. The standard assumption is vanishing anisotropic stress, i.e., $\sigma_{\rm{DE}}=0$. Lastly, the adiabatic sound speed is defined as
\begin{equation}
c_{\rm{a}}^2 \equiv \frac{{p}_{\rm{DE}}^\prime}{{\rho}_{\rm{DE}}^\prime} \,.
\end{equation}

This fluid parameterization is physically intuitive and adequately describes a range of DE models; however, in scenarios where $1+w$ crosses zero (dubbed phantom crossing), Eq.~\eqref{eq:fluid-euler} becomes ill-behaved. In such cases, an alternative prescription is required.

\subsubsection{Parametrized post-Friedmann dark energy}

The PPF formalism was developed to address the shortcomings of the fluid approach near the phantom crossing. In the PPF approach, DE perturbations are recast in terms of a new variable, $\Gamma$, defined in the DE rest frame by
\begin{equation}
k^2 \Gamma \equiv -4 \pi G \, a^2 \, \delta \rho_{\rm{DE}}^{\rm{rest}} \,.
\end{equation}
In the conformal Newtonian gauge, the Poisson equation is written as
\begin{equation}
k^2 \phi = -4 \pi G \, a^2 \left[ \delta \rho_{\rm{total}}^{\rm{N}} - 3 \mathcal{H} \left( \rho_{\rm{total}} + p_{\rm{total}} \right) \frac{\theta_{\rm{total}}^{\rm{N}}}{k^2} \right] \,,\label{eq:ppf_total}
\end{equation}
where the subscript ``total'' tautologically indicates the contributions from all species, not only from DE, and the superscript N denotes the Newtonian gauge. One can split the contributions of DE from the other species using that
\begin{equation}
(1+w_{\rm{total}})\,\theta_{\rm{total}} = \sum_i (1+w_i)\,\frac{\rho_i}{\rho_{\rm{total}}}\,\theta_i \,,
\end{equation}
where
\begin{equation}
w_{\text{total}} = \frac{\sum_i w_i\,\rho_i}{\sum_i \rho_i} \quad \text{and} \quad \rho_{\text{total}} = \sum_i \rho_i \,,
\end{equation}
and the summation is made over DE and the other components. Therefore, Eq.~\eqref{eq:ppf_total} can be rewritten as
\begin{equation}
k^2 \phi = -4 \pi G \, a^2 \left[ \delta \rho_{\rm{t}}^{\rm{N}} - 3 \mathcal{H} \left( \rho_{\rm{t}} + p_{\rm{t}} \right) \frac{\theta_{\rm{t}}^{\rm{N}}}{k^2} \right] + k^2 \Gamma \,,\label{eq:ppf_split}
\end{equation}
where we use the same notation as~\citet{Dakin:2019vnj} and the subscript ``t'' represents the contributions of all components \emph{but} DE. Imposing the physical requirements of energy conservation on super-horizon scales and recovering the standard Newtonian Poisson equation on small scales implies that $\Gamma$ satisfies
\begin{equation}
{\Gamma}^\prime = \mathcal{H} \left( S - \Gamma \right) \,,
\end{equation}
with
\begin{equation}
S \equiv \frac{4 \pi G \, a^2}{\mathcal{H}} \left( \rho_{\rm{DE}} + p_{\rm{DE}} \right) \frac{\theta_{\rm{t}}^{\rm{N}}}{k^2} \,.
\end{equation}
To interpolate between the super-horizon regime and the small-scale limit, an effective parameter $c_\Gamma$ is introduced so that the evolution of $\Gamma$ is governed by
\begin{equation}
{\Gamma}^\prime = \mathcal{H} \left[ \frac{S}{1 + c_\Gamma^2 k^2 / \mathcal{H}^2} - \Gamma \left( 1 + \frac{c_\Gamma^2 k^2}{\mathcal{H}^2} \right) \right] \,.
\label{eq:PPF-Gamma}
\end{equation}
A value of $c_\Gamma \sim 0.4\, c_{\rm{s}}$ reproduces the behavior of quintessence models and is widely adopted. This PPF approach provides a robust description of DE perturbations that remain well-behaved even when the fluid description fails.

\citet{Dakin:2019vnj} present a careful study of DE perturbations in $N$-body simulations by comparing the fluid and the PPF implementations within the \concept\ code~\citep{Dakin:2017idt,Dakin:2019vnj,Dakin:2021ivb}.\footnote{\url{https://github.com/jmd-dk/concept/}} They show that both methods yield similar results on scales $k \gtrsim 10^{-3} h/{\rm Mpc}$. In contrast, the methods deviate significantly on larger scales. In addition, for a sound speed equal to unity, both parameterizations agree with Newtonian predictions at $k \gtrsim 10^{-2} h/{\rm Mpc}$, while on larger scales, general relativistic corrections due to DE perturbations cause deviations of tens of percent. Furthermore, percent level differences in the matter power-spectrum at $k \sim 10^{-2} h/{\rm Mpc}$ are only achieved for $c_{\rm s}^2\lesssim10^{-2}$.

\subsection{The halo mass function}
\label{sec:hmf}

The HMF describes the comoving number density of matter halos as a function of mass and redshift. It is conventionally written in differential form as
\begin{equation}
\frac{dn}{dM}\,dM = \frac{\rho_{\rm m}}{M}\,\nu\,f(\nu)\,d\ln \nu\,,
\label{eq:hmf_new}
\end{equation}
where \(\rho_{\rm m}\) represents the mean comoving matter density of the Universe, and \(\nu\) is the peak height parameter. This parameter is defined as
\begin{equation}
\nu = \frac{\delta_{\rm c}}{\sigma(M,z)}\,,
\label{eq:nu_new}
\end{equation}
with \(\delta_{\rm c}\) being the critical linear overdensity for spherical collapse extrapolated to redshift $z$~\citep{Kitayama:1996ne} and \(\sigma(M,z)\) the mass variance. The mass variance is computed from the linear matter power spectrum \(P_{\rm m}(k,z)\) through
\begin{equation}
\sigma^2(M,z) = \frac{1}{2\pi^2} \int_0^\infty dk\,k^2\,P_{\rm m}(k,z)\,W^2(kR)\,,
\label{eq:sigma_new}
\end{equation}
where \(R\) is the Lagrangian radius corresponding to mass \(M\) (i.e., \(R = \left[3M/(4\pi\rho_{\rm m})\right]^{1/3}\)) and \(W(kR)\) is the Fourier transform of a top-hat window function.

In \citet{Euclid:2022dbc}, we demonstrated that the multiplicity function $\nu f(\nu)$ given by
\begin{equation}
\nu f(\nu) = A(p, q) \sqrt{\frac{2 a \nu^2}{\pi}} \exp\left( -\frac{a \nu^2}{2} \right) \left[ 1 + \frac{1}{(a \nu^2)^p} \right] (\sqrt{a} \nu)^{q - 1}\,,
\label{eq:mult}
\end{equation}
provides flexibility to reproduce the outcomes of several scale-free simulations accurately. In that work, the challenge of capturing the non-universality of the HMF was reduced mainly to describing the cosmological evolution of the free parameters that appear in Eq.~\eqref{eq:mult}. It was shown that both the background evolution and the power spectrum shape impact the HMF universality. Their cosmological dependency was modeled as
\begin{align}
a &= a_R \, \Omega_{\rm m}^{a_z}(z)\,, \label{eq:a_param} \\
a_R &= a_1 + a_2 \left( \frac{\de \ln \sigma}{\de \ln R} + 0.6125 \right)^2\,, \label{eq:aR_param} \\
p &= p_1 + p_2 \left( \frac{\de \ln \sigma}{\de \ln R} + 0.5 \right)\,, \label{eq:p_param} \\
q &= q_R \, \Omega_{\rm m}^{q_z}(z)\,, \label{eq:q_param}
\\
q_R &= q_1 + q_2 \left( \frac{\de \ln \sigma}{\de \ln R} + 0.5 \right)\,,\label{eq:qR_param}
\end{align}
and we refer the reader to~\citet{Euclid:2022dbc} for the values of the free parameters.

For the DUCA simulation suite, which spans a considerably broader dynamic range, we found that the previous baseline function lacked the flexibility to capture the simulation data fully and proposed in~Paper~I the following parametrization for the HMF on a scenario with dynamic DE
\begin{equation}
    \nu f(\nu) = A(p, q, r)\,\sqrt{\frac{2\nu_*^2}{\pi}} \,\exp\left(-\frac{\nu_*^2}{2}\right)\,\left\{ \nu_*^r + \left( \frac{\nu_p}{\nu_*} \right)^{2p} \right\}\,\nu_*^{\,q-1}\,,
    \label{eq:new_mult_spin}
\end{equation}
where we define \(\nu_*=\sqrt{a}\,\nu\). In this formulation, the normalization constant \(A\) is determined by
\begin{equation}   
\begin{split}
    \left(\frac{A}{\Xi}\right)^{-1} =&\; -2^{p+\frac{r}{2}}\,q\,\Gamma\!\left(\frac{q}{2}+\frac{r}{2}+1\right) + 2^{p+\frac{r}{2}+1}\,p\,\Gamma\!\left(\frac{q}{2}+\frac{r}{2}+1\right)\\[1ex]
    &\; -\,q\,\nu_p^{2p}\,\Gamma\!\left(-p+\frac{q}{2}+1\right) - r\,\nu_p^{2p}\,\Gamma\!\left(-p+\frac{q}{2}+1\right)\,,
\end{split}
\end{equation}
with
\begin{equation}
    \Xi = \frac{\sqrt{\pi}\,(2p-q)(q+r)}{2^{-p+\frac{q}{2}+\frac{1}{2}}}\,.
\end{equation}
In the limits \(r\to 0\) and \(\nu_p\to 1\), Eq.~\eqref{eq:new_mult_spin} naturally reverts to the multiplicity function given in Eq.~\eqref{eq:mult}. The dependence on DE is introduced by allowing the parameters of the multiplicity function, collectively denoted as $x_i\in\{a, p, q, r\}$, to vary with cosmology while $\nu_p$ is cosmology independent. The cosmological dependency of the parameters $x_i$ is modeled according to
\begin{equation}
x_i = x_{\Lambda,i}\left\{ 1 + \frac{3\,\alpha_i}{2}\,\left[w_{\rm DE}(z_{\rm ta})+1\right]\,\Omega_{\rm DE}(z_{\rm ta}) \right\}\,,
\label{eq:dedependency}
\end{equation}
where $x_{\Lambda,i}$ is given by Eqs.~\eqref{eq:a_param}--\eqref{eq:qR_param} for $x_i\in\{a, p, q\}$ while
\begin{equation}
r = r_1 + r_2 \left( \frac{\de \ln \sigma}{\de \ln R} + 0.5 \right)\,, \label{eq:R_param}
\end{equation}
In Eq.~\eqref{eq:dedependency}, $z_{\rm ta}$ is the turnaround redshift at which a spherical top-hat perturbation, collapsing at redshift $z$, reaches its maximum expansion. The free parameters $\alpha_i$ modulate the impact of the evolving DE component on the HMF. This formulation links the modifications in the multiplicity function to the epoch when the effects of DE and gravitational collapse are in balance. We adopt the HMF presented in Paper~I as our baseline model and, for conciseness, we refer to Paper~I for further details on the implementation and values of the free parameters in Eq.~\eqref{eq:new_mult_spin}.

\section{Methodology}
\label{sec:methodology}

In this section, we present our methodology. We introduce the simulations produced in~Sect.~\ref{sec:sims}, the halo catalogs in Sect.~\ref{sec:halofinder}, and present the model calibration set-up in Sect.~\ref{sec:cal}.

\subsection{Simulations}
\label{sec:sims}

We have performed 16 cosmological simulations using the \concept\ code, employing the same integration accuracy settings as in the DUCA simulation set presented in Paper~I. All runs were carried out in a cubic volume with a comoving side length of $2\,h^{-1}\,\mathrm{Gpc}$ and evolved with $2048^3$ particles, ensuring a particle mass of roughly $8\times 10^{10}\,h^{-1}\,M_\odot$ which provides the resolution needed for our analyses in the regime of objects more massive than $2\times 10^{13}\,h^{-1}\,M_\odot$. The initial conditions were generated on the fly using third-order Lagrangian perturbation theory (3LPT) at a starting redshift of $z=24$.

The background cosmology, with the exception of the DE parameters, is kept fixed in all simulations with the following parameters: a Hubble constant of $H_0 = 67\,\mathrm{km\,s^{-1}\,Mpc^{-1}}$, a baryon density parameter $\Omega_{\rm b} = 0.049$, and a CDM density parameter $\Omega_{\rm cdm} = 0.27$.  The primordial power spectrum is specified by an amplitude $A_{\rm s} = 2.1\times10^{-9}$, a spectral index $n_{\rm s} = 0.96$, zero running ($\alpha_{\rm s} = 0$), and a pivot scale of $k_{\rm pivot} = 0.05\,\mathrm{Mpc}^{-1}$.

We vary the DE parameters across the simulation suite to investigate the impact of DE dynamics. In particular, the DE sound speed squared, $c_{\rm s}^2$, is set to one of four values ($10^{-7}$, $10^{-5}$, $10^{-3}$, or 1), thereby spanning a wide range of clustering behaviors. For each choice of $c_{\rm s}^2$, we explore three combinations of the DE EOS parameters by varying $w_0$ and $w_a$. Specifically, $w_0$ takes the values $-1.0$, $-0.9$, and $-0.8$, while the corresponding $w_a$ values are $0.3$, $0.2$, and $0.1$, respectively. This systematic variation results in 12 simulations using the PPF prescription, which enables a comprehensive exploration of the phenomenology associated with dynamical DE. 

Table~\ref{tab:simulations} summarizes our simulation suite's DE parameter combinations, including four additional simulations: simulations~\#3 and~\#11 have also been run using the fluid description and its phantom twin -- a simulation with DE EOS reflected with respect to $-1$.

We selected the sound speed values in our simulations to examine various CDE levels while working within the constraints of the linear‐species implementation in \concept.  These constraints impose a lower bound on $c_{\rm s}$ as the realization method cannot accurately simulate values that are too low.  In \citet{Hassani:2019lmy}, the linear‐realization scheme was compared with an alternative EFT approach for DE perturbations.  The resulting difference in the non‐linear matter power spectrum is below 1 percent for $c_{\rm s}^2=10^{-7}$ at $z=0$ and decreases further at higher redshift.  While this finding motivates our choice of $c_{\rm s}$, it is essential to remember that our implementation remains at the level of linear DE perturbations, capturing the impact of these perturbations on the non-linear collapse of dark matter but, by construction, it excludes any backreaction of non-linear matter clustering on the DE field. Furthermore, the DE equation of state can also influence the CDE non‐linear behavior~\citep{Batista:2022ixz,Nouri-Zonoz:2024dph}. Therefore, our lowest‐$c_{\rm s}$ results should be regarded as a baseline, pending more advanced treatments of non‐linear CDE.

\begin{table}[ht]
    \centering
    \caption{DE parameter combinations for the simulation suite.}
    \label{tab:simulations}
    \begin{tabular}{ccccccc}
        \hline\hline
        Simulation & $c_{\rm s}^2$ & $w_0$ & $w_a$ & PPF & Fluid & Phantom \\
          \#      &               &       &       &     &       & Twin \\
        \hline
        1  & $10^{-7}$ & $-1.0$ & $0.3$ & \cmark & \xmark & \xmark\\
        2  & $10^{-7}$ & $-0.9$ & $0.2$ & \cmark & \xmark & \xmark\\
        3  & $10^{-7}$ & $-0.8$ & $0.1$ & \cmark & \cmark & \cmark\\
        4  & $10^{-5}$ & $-1.0$ & $0.3$ & \cmark & \xmark & \xmark\\
        5  & $10^{-5}$ & $-0.9$ & $0.2$ & \cmark & \xmark & \xmark\\
        6  & $10^{-5}$ & $-0.8$ & $0.1$ & \cmark & \xmark & \xmark\\
        7  & $10^{-3}$ & $-1.0$ & $0.3$ & \cmark & \xmark & \xmark\\
        8  & $10^{-3}$ & $-0.9$ & $0.2$ & \cmark & \xmark & \xmark\\
        9  & $10^{-3}$ & $-0.8$ & $0.1$ & \cmark & \xmark & \xmark\\
        10 & $1$       & $-1.0$ & $0.3$ & \cmark & \xmark & \xmark\\
        11 & $1$       & $-0.9$ & $0.2$ & \cmark & \cmark & \cmark\\
        12 & $1$       & $-0.8$ & $0.1$ & \cmark & \xmark & \xmark\\
        \hline
    \end{tabular}
\tablefoot{The background cosmology is kept fixed in all simulations: $H_0 = 67\,\mathrm{km\,s^{-1}\,Mpc^{-1}}$, $\Omega_{\rm b} = 0.049$, $\Omega_{\rm cdm} = 0.27$, $A_{\rm s} = 2.1\times10^{-9}$, $n_{\rm s} = 0.96$, $\alpha_{\rm s} = 0$, and $k_{\rm pivot} = 0.05\,\mathrm{Mpc}^{-1}$. The simulations start at $a_{\rm begin} = 0.04$ from 3LPT initial conditions. Phantom twin simulations have a DE EOS, which is reflected with respect to $w=-1$ compared to the reference value.}
\end{table}

\subsection{Halo identification}
\label{sec:halofinder}

In our study, we make use of the \rockstar\footnote{\url{https://bitbucket.org/gfcstanford/rockstar}} halo finder in combination with the \consistent\ Trees algorithm.\footnote{\url{https://bitbucket.org/pbehroozi/consistent-trees}} \rockstar\ is an advanced phase-space halo finder that employs both spatial and velocity information to detect dark matter halos \citep{Behroozi:2011ju}.

The procedure commences by dividing the simulation volume into Friends-of-Friends (FoF) groups using a relatively large linking length of 0.28 (expressed in units of the mean interparticle separation), which exceeds the conventional value of approximately 0.20 typically used in other FoF algorithms. This initial grouping facilitates efficient parallel processing. Within each FoF group, \rockstar\ performs an adaptive hierarchical refinement in six-dimensional phase space (three dimensions for position and three for velocity), resulting in a nested hierarchy of subgroups. This approach permits the precise identification of halos and their substructures, even in regions with high particle density.

To improve both temporal consistency and the accuracy of halo properties, we process the \rockstar\ outputs with the \consistent\ Trees algorithm \citep{Behroozi:2011js}. This algorithm tracks halos across successive simulation snapshots, ensuring that properties such as halo masses and positions evolve smoothly over time. Such dynamic tracking is essential for reducing scatter and enhancing the reliability of the halo catalogs in our analysis.

We adopt the spherical overdensity (SO) criterion for defining halos, whereby the mass of a halo is determined by the mean density within a sphere that is equal to the virial overdensity $\Delta_{\rm vir}(z)$.  \rockstar\ uses the fitting formula of \citet{Bryan_1998} for $\Delta_{\rm vir}$ and considers only the contribution of matter.  In the case of CDE, including the DE contribution within the overdensity radius could, in principle, improve the universality of the HMF, as suitable choices of the collapse threshold are known to preserve universality in homogeneous DE scenarios~\citep{Diemer:2020rgd}.  However, a self-consistent CDE spherical collapse calculation would introduce significant computational overhead and complicate the HMF evaluation, and we therefore retain the standard matter‐only~\citet{Bryan_1998} fit as implemented in \rockstar.

\rockstar\ defines the SO center as the average positions of the particles belonging to the innermost subgroup in the hierarchy. All particles within the virial radius are included in the halo mass, irrespective of their gravitational binding.

Finally, the halo catalogs generated by the combined use of \rockstar\ and \consistent\ Trees are post-processed, counting the number of objects as a function of halo mass in logarithmically spaced bins based on the number of particles per halo. This binning strategy minimizes the effects of mass discretization and ensures a statistically robust sampling across the halo mass range. Only halos containing more than 300 particles are considered. This selection cut in mass is sufficient to ensure sub-percent agreement with higher-resolution simulations~\citep[see, for instance,][]{Euclid:2022dbc}.

\subsection{\label{sec:cal}Model calibration}

We adopt a Bayesian approach to calibrate the HMF model. The calibration procedure follows that of~\citet{Euclid:2022dbc} and in Paper~I; the reader is referred to those works for a complete discussion. Here, we summarize the main aspects.

The calibration fits the theoretical prediction as a function of the HMF parameters $\pmb{\theta}$ 
\begin{equation}
N_i^{\rm th} = N_i(\pmb{\theta}, z)\,,    
\end{equation}
obtained by integrating the HMF model over the mass bin and multiplying by the simulation volume, to the halo counts \(N_i^{\rm sim}\) extracted from our simulations (see Sect.~\ref{sec:halofinder}). The number of halos in each mass bin is assumed to follow a Poisson distribution as motivated by the Press--Schechter formalism~\citep{Press:1973iz}. The Poisson log-likelihood is given by
\begin{equation}
\ln \mathcal{L}(N_i^{\rm sim} \mid \pmb{\theta}, z) = N_i^{\rm sim} \ln N_i^{\rm th} - N_i^{\rm th} - \ln\bigl(N_i^{\rm sim}!\bigr)\,.
\label{eq:poisson_likelihood}
\end{equation}
To account for systematic effects and numerical uncertainties, a composite likelihood is adopted. In bins with a large number of halos (i.e., \(N_i^{\rm sim} > 25\)), the Poisson likelihood is approximated by a Gaussian distribution~\citep{Euclid:2022dbc}. The composite log-likelihood is expressed as
\begin{equation}
\ln \mathcal{L}(N_i^{\rm sim} \mid \pmb{\theta}, z) =
\begin{cases}
\begin{aligned}
&N_i^{\rm sim} \ln N_i^{\rm th} - N_i^{\rm th} \\
&- \ln\bigl(N_i^{\rm sim}!\bigr)
\end{aligned} & \text{if } N_i^{\rm sim} \leq 25\,, \\[5ex]
\begin{aligned}
&- \dfrac{1}{2} \left( \dfrac{N_i^{\rm sim} - N_i^{\rm th}}{\sigma_i} \right)^2 \\
&- \dfrac{1}{2} \ln\bigl(2\pi \sigma_i^2\bigr)
\end{aligned} & \text{if } N_i^{\rm sim} > 25\,,
\end{cases}
\label{eq:composite_likelihood}
\end{equation}
with the standard deviation defined as
\begin{equation}
\sigma_i^2 = N_i^{\rm th}\,\left(1 + N_i^{\rm th}\sigma_{\rm sys}^2\right)\,.
\label{eq:likelihood_sigma}
\end{equation}
In this formulation, \(\sigma_{\rm sys}\) represents the additional variance component to account for systematic uncertainties and is fixed to the best fit presented in Paper~I, namely $8.82\times10^{-2}$. This correction is not included in the Poissonian part of the likelihood.

The overall log-likelihood is computed by summing over all mass bins, redshifts, and simulation outputs
\begin{equation}
\ln \mathcal{L}_{\rm total}(\pmb{\theta}) = \sum_{s} \sum_{z} \sum_{i} \ln \mathcal{L}(N_{i,z,s}^{\rm sim} \mid \pmb{\theta}, z)\,.
\label{eq:total_likelihood}
\end{equation}
In the above equation, the index \(s\) labels the simulation outputs. Although outputs from the same simulation at different redshifts are not strictly independent, this issue is mitigated by selecting snapshots separated by intervals exceeding the typical dynamical time of galaxy clusters (approximately \(1.7\) Gyr) in accordance with~\citet{Bocquet:2015pva}.

The best-fit value for $\pmb{\theta}$ is determined by the global maximum identified with the \textsc{Dual Annealing} algorithm~\citep{1988JSP....52..479T,1996PhyA..233..395T,1997PhLA..233..216X,PhysRevE.62.4473} as implemented in \textsc{scipy}~\citep{Virtanen:2019joe}.

\section{Results}
\label{sec:results}

\subsection{Impact of variable sound speed}
\label{sec:cs_effects}

\begin{figure*}
    \centering
    \includegraphics[width=0.99\textwidth]{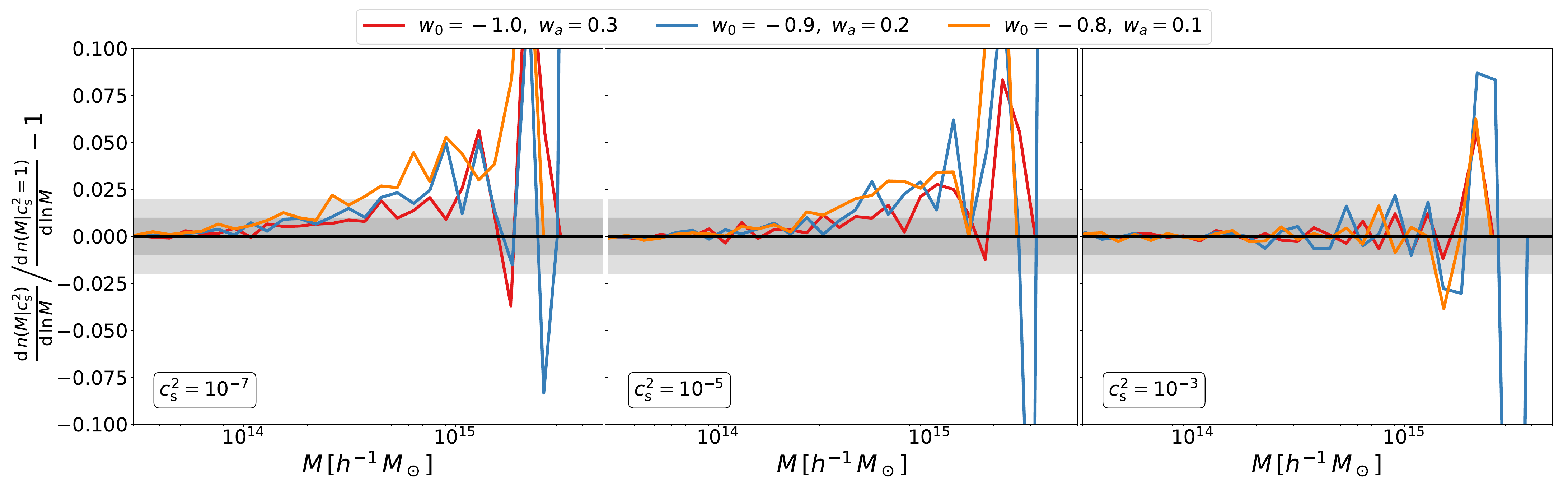}
    \caption{Relative differences in HMF at redshift zero for varying DE sound speeds $c_{\rm s}^2$, normalized to the reference case $c_{\rm s}^2=1$. Different colors correspond to different DE EOS parameters, while different panels correspond to different sound speeds. From left to right, $c_{\rm s}^2$ is $10^{-7}$, $10^{-5}$, and $10^{-3}$. Gray bands mark $\pm 1 \%$ (dark) and $\pm 2 \%$ (light) differences. 
}
    \label{fig:cs2}
\end{figure*}

\begin{figure}
    \centering
    \includegraphics[width=0.99\columnwidth]{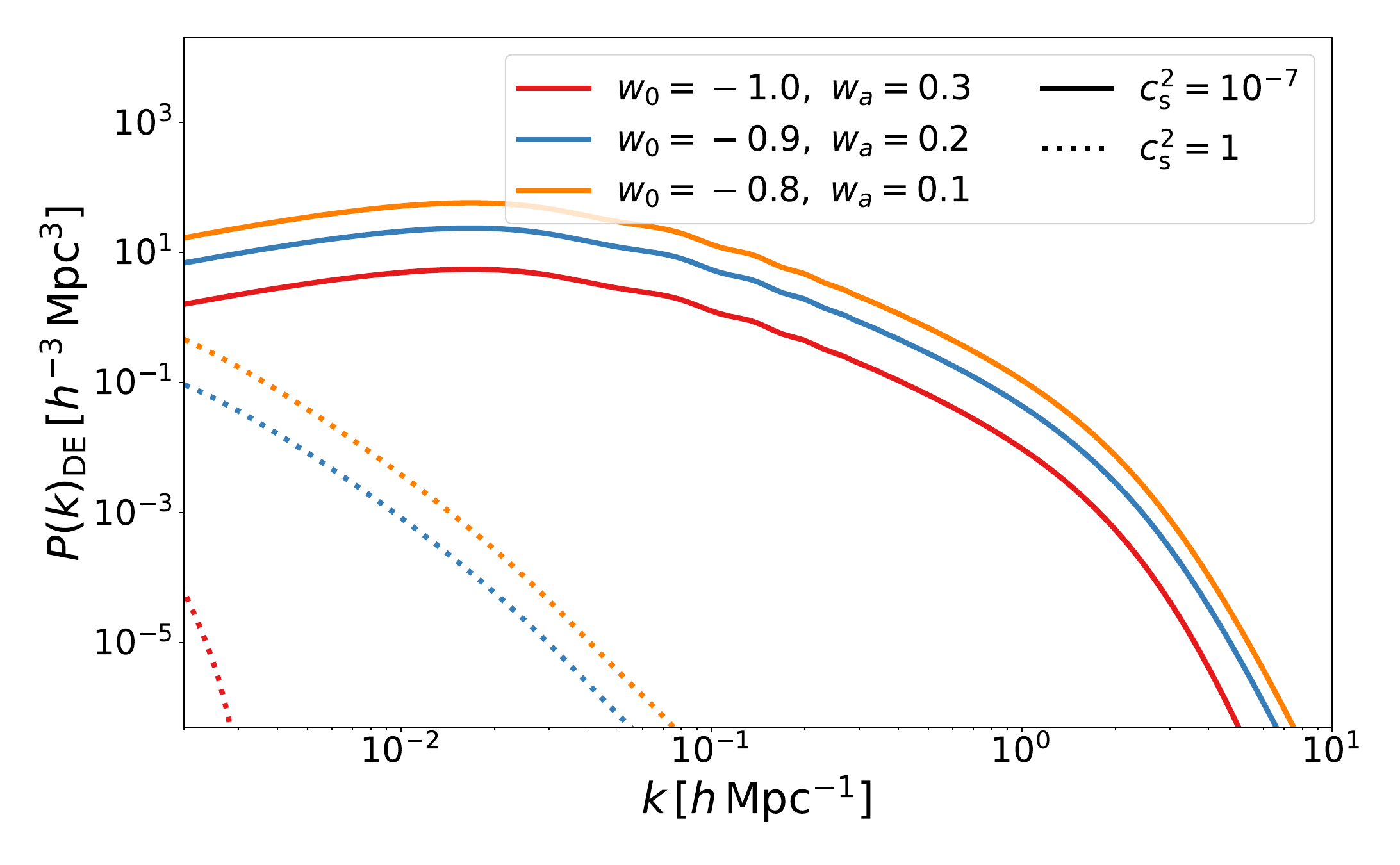}
    \caption{DE power spectrum at redshift zero and sound speeds $c_{\rm s}^2=10^{-7}$ and $c_{\rm s}^2=10^{-7}$ for the different DE EOS parameters.
}
    \label{fig:pk_de}
\end{figure}

In Fig.~\ref{fig:cs2}, we present the relative differences in HMF at redshift zero, when the CDE impact is the strongest, for varying DE sound speeds $c_{\rm s}^2$, normalized to the reference case $c_{\rm s}^2=1$. Different colors correspond to different sets of parameters for the DE EOS (see also Table~\ref{tab:simulations}), while different panels correspond to different sound speeds. From left to right, $c_{\rm s}^2$ is $10^{-7}$, $10^{-5}$, and $10^{-3}$. 

Not surprisingly, the impact of CDE on the HMF is more substantial the lower the sound speed, as, in this case, its clustering is higher. While in the case for $c_{\rm s}^2=10^{-3}$ we only observe an increasing scatter of the number counts at higher masses, for  $c_{\rm s}^2=10^{-5}$ and $c_{\rm s}^2=10^{-7}$ we observe a slight boost in the number of objects of the order of few percent for the most massive objects. 

From Fig.~\ref{fig:cs2}, we also observe that the larger the deviation from $w=-1$, the higher the impact; again, as in this condition, DE perturbations are more important. To verify that, we present in Fig.~\ref{fig:pk_de} the DE power spectrum computed using \concept. Therefore, the higher the clustering in DE, the higher the impact on the HMF.

\subsection{Extended model}
\label{sec:model}

\begin{figure}
    \centering
    \includegraphics[width=0.99\columnwidth]{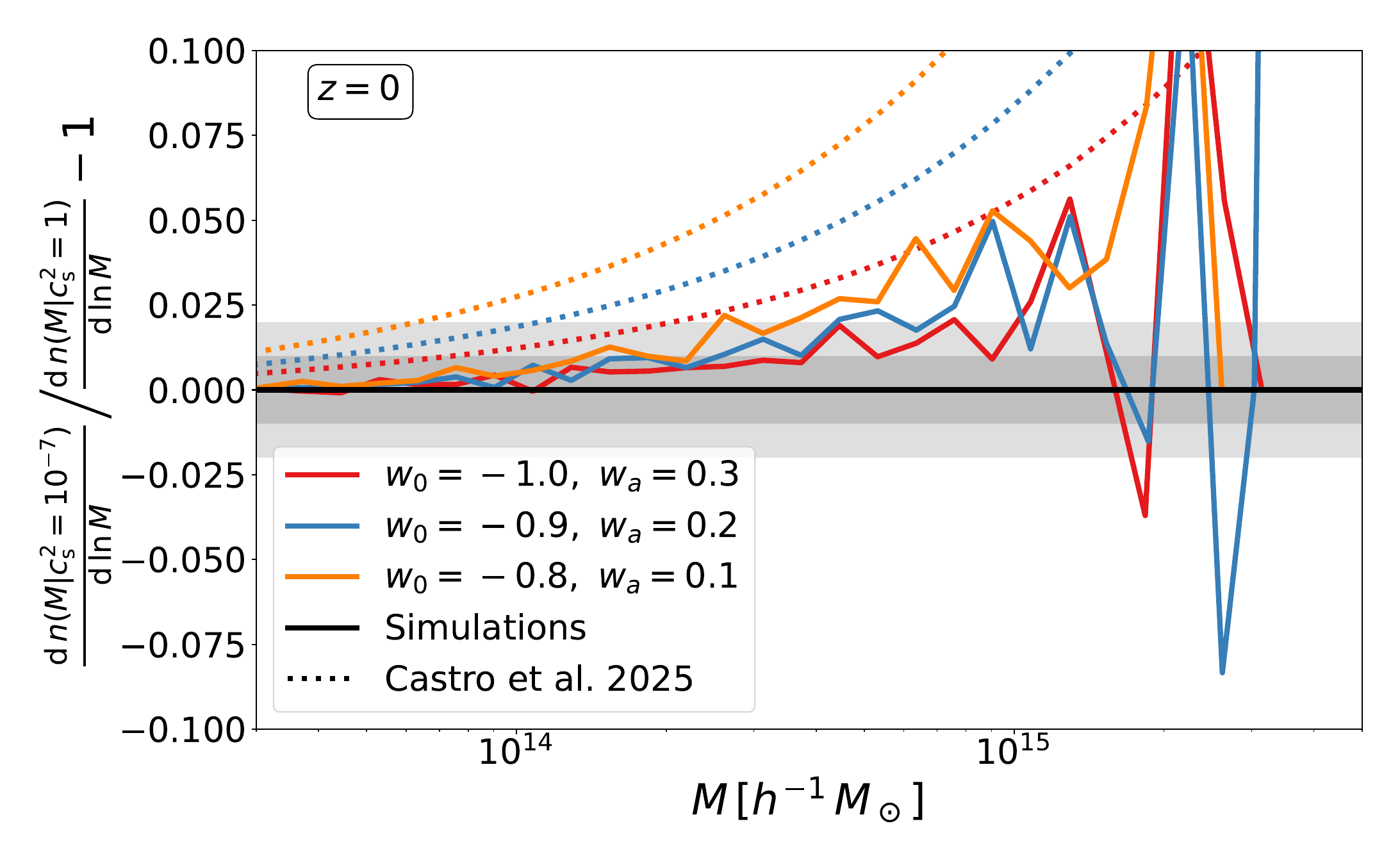}
    \caption{Relative differences in the halo mass function (HMF) at $z = 0$ for the case where $c_{\rm s}^2 = 10^{-7}$, compared to the reference case $c_{\rm s}^2 = 1$, for different DE EOS parameters (filled lines in different colors). The figure also shows the prediction of this quantity from our baseline model, presented in Paper~I, as indicated by the dotted lines.}
    \label{fig:castro25}
\end{figure}

\begin{figure*}
    \centering
    \includegraphics[width=0.99\textwidth]{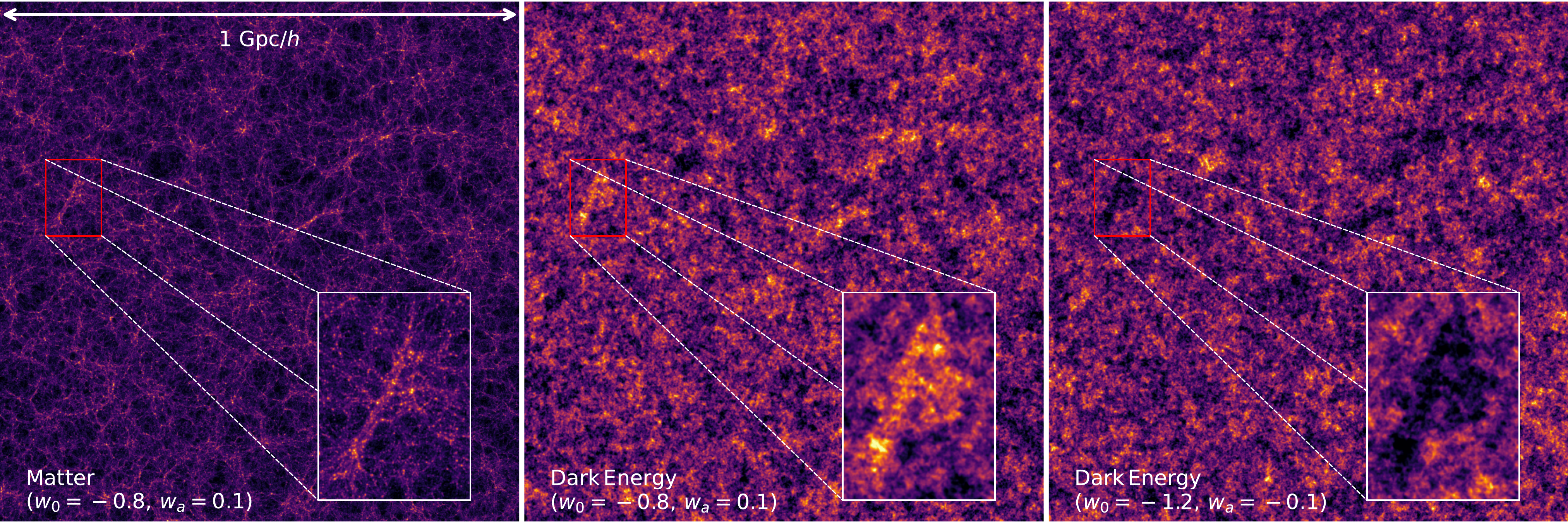}
    \caption{The density fields in a comoving volume of size 1\,Gpc$/h$ at $z=0$. 
    The left panel shows the matter distribution for the simulation~\#3, i.e., $w_0 = -0.8,\, w_a = 0.1$. 
    The middle and right panels show the DE distributions for the same simulation and its phantom twin $(w_0 = -1.2,\, w_a = -0.1)$, respectively. The region inside the red rectangles is zoomed and displayed in an inset at the lower right. 
    }
    \label{fig:densityfield}
\end{figure*}

In Fig.\ref{fig:castro25}, we compare the relative differences in the HMF at $z=0$ for $c_{\rm s}^2=10^{-7}$ with respect to the reference case $c_{\rm s}^2=1$, alongside the prediction of our baseline model from Paper~I. The baseline model accounts for the impact of CDE on clustering via Eq.~\eqref{eq:sigma_new}, where $P_{\rm m}$ is computed directly from simulations with $c_{\rm s}^2=10^{-7}$. Although it qualitatively reproduces the simulation results, it overestimates the effect by a factor of a few.

In order to understand the reason for the mismatch between the baseline model and the simulations we have to keep in mind the key role played by the mapping between linear theory and nonlinear clustering within the HMF formalism. Relativistic species, which couple to matter only through gravity, cluster significantly less than matter. As a result, linear theory overestimates the actual impact that CDE has on nonlinear matter clustering. Consequently, using the linear matter power spectrum for $c_{\rm s}^2<1$ to compute the mass variance in Eq.~\eqref{eq:sigma_new} and to predict the number of collapsed structures leads to an overestimate of the effect of CDE. A similar conclusion was shown by~\citet{Nouri-Zonoz:2024dph} studying the linear and non-linear coupling between DE and matter perturbations.

To reconcile this issue with the sensitivity of the HMF to CDE clustering, we propose the following empirical modification to the computation of the peak height:
\begin{equation}
\nu(M, z|c{_{\rm s}^2<1}) \rightarrow \frac{\nu(M, z|c{_{\rm s}^2=1})}{1 + \lambda\,\mathrm{sgn}\Bigl(1+w(z_{\rm ta})\Bigr)\,\frac{\Omega_{\rm DE}(z_{\rm ta})\,\sigma_{\rm DE}(R_{\rm ta}, z_{\rm ta}|c{_{\rm s}^2<1})}{\Omega_{\rm m}(z_{\rm ta})\,\sigma_{\rm m}(R, z_{\rm ta}|c{_{\rm s}^2<1})}\,\frac{\delta_{\rm ta}^{\rm EdS}}{\Delta_{\rm ta}^{\rm EdS}}}\,.
\label{eq:sigma_mod}
\end{equation}
Here $\sigma_{\rm DE}$ and $\sigma_{\rm m}$ are the variance of DE and non-relativistic matter perturbations, respectively, $\mathrm{sgn}$ is the sign function, and the subscript “ta” indicates that the quantities are computed at the turnaround. As in~Paper~I, the Einstein--de Sitter (EdS) values are used for the turnaround redshift as well as for the linear and non-linear density contrasts, $\delta_{\rm ta}^{\rm EdS}$ and $\Delta_{\rm ta}^{\rm EdS}$, respectively. Numerically, they are
\begin{align}
z_{\rm ta}&=\frac{1+z}{(1/2)^{2/3}}-1\,,\\
\delta_{\rm ta}^{\rm EdS}&=\frac{9\,\pi^2}{16}\,,\\
\Delta_{\rm ta}^{\rm EdS}&=\frac{3}{5}\left(\frac{3\,\pi}{4}\right)^{2/3}\,.
\end{align}

The following considerations motivate the terms in Eq.~\eqref{eq:sigma_mod}. The sign function, $\mathrm{sgn}\Bigl(1+w(z_{\rm ta})\Bigr)$, is introduced because when the DE EOS crosses $-1$, the pressure contribution overtakes the density contribution in affecting the gravitational potential; this leads to a de-boost in matter clustering. This is illustrated in Fig.~\ref{fig:densityfield}, where we note that a region of overdensity of matter corresponds to an overdensity of DE for the simulation~\#3 ~\#3, but to an underdensity of DE on its phantom-twin counterpart, in agreement with the results of~\citet{Batista:2022ixz}. In addition, the ratio of the density parameters, $\Omega_{\rm DE}(z_{\rm ta})/\Omega_{\rm m}(z_{\rm ta})$, is included to weight the DE perturbations relative to those of matter, as the gravitational impact is governed by the absolute density perturbations—as reflected in the Euler equation in the $N$-body gauge—rather than by the density contrast alone. The ratio of variances, $\sigma_{\rm DE}(R_{\rm ta}, z_{\rm ta}|c{_{\rm s}^2<1})/\sigma_{\rm m}(R, z_{\rm ta}|c{_{\rm s}^2<1})$, quantifies the relative amplitude of DE fluctuations compared to matter fluctuations. In this context, evaluating the variance for DE fluctuations at the turnaround scale is crucial because the late onset of DE clustering implies that, when DE becomes dynamically important, the collapsing matter is already confined to a smaller volume than the original Lagrangian patch $R$. As CDE perturbations are well described by linear theory virtually at all scales and epochs, the difference between Eulerian and Lagrangian coordinates of a CDE fluid element is always small. Therefore, to assess the impact that CDE has on the DM collapse, one has to compare the DE fluctuations at the scale when DE is dynamically and kinematically relevant for the collapse -- in our model, the turnaround radius -- to the fluctuations of matter perturbations linearly extrapolated to the epoch of turnaround.
Note also, from Eq.~\eqref{eq:fluid-continuity}, that DE perturbations go to zero when the EOS parameter approaches $-1$.
Finally, the factor $\delta_{\rm ta}^{\rm EdS}/\Delta_{\rm ta}^{\rm EdS}$ is introduced to rescale the matter variance computed from the linear power spectrum, correcting for the mismatch that arises because DE perturbations remain in the linear regime at the turnaround. In contrast, matter perturbations have already entered the non-linear regime. Finally, the empirical constant $\lambda$ calibrates the overall strength of the DE clustering effect. Together, these modifications yield a peak height that more accurately reflects the interplay between DE clustering and matter perturbations in structure formation.

\subsection{Model calibration and validation}
\label{sec:robustness}

\begin{figure*}
    \centering
    \includegraphics[width=0.99\textwidth]{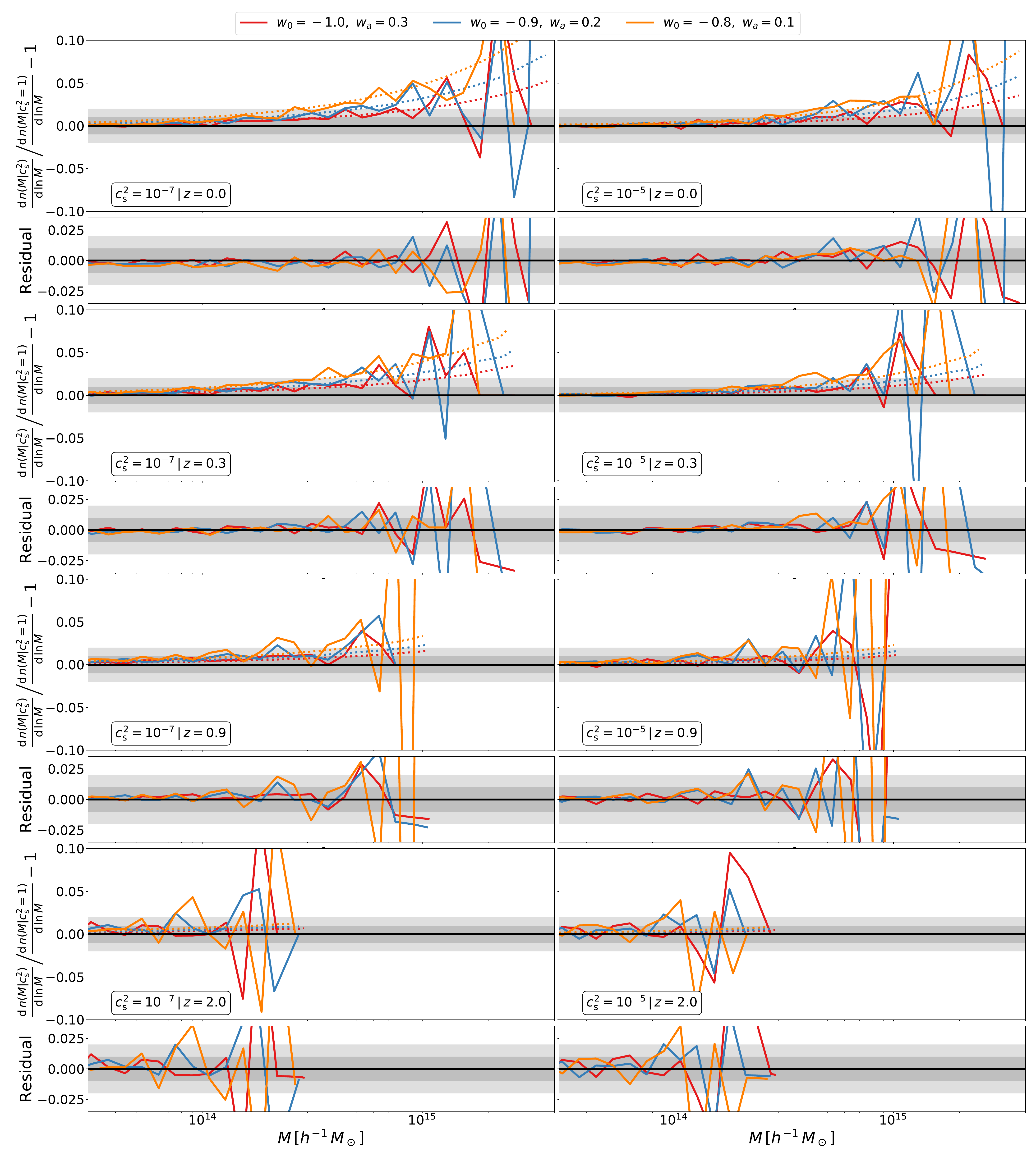}
    \caption{Comparison between the model calibrated on the Simulations 1--12 using the PPF description presented in Table~\ref{tab:simulations} (dotted) and the simulations for $c_{\rm s}^2 \in \{10^{-7}, 10^{-5}\}$ (solid lines). Different DE EOS are presented in different colors, and the rows present the comparison and respective relative residuals at the selected redshifts $z\in\{0.0, 0.3, 0.9, 2.0\}$.}
    \label{fig:newmodel}
\end{figure*}

\begin{figure*}
    \centering
    \includegraphics[width=0.99\textwidth]{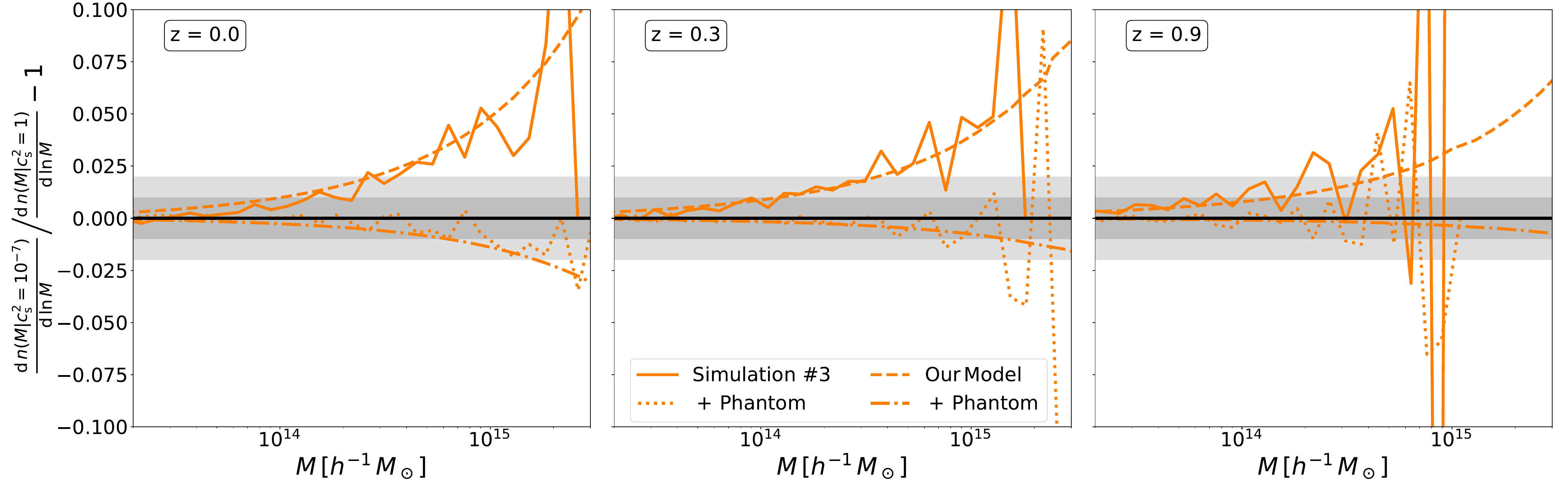}
    \caption{Comparison between the best fit model prediction for the cosmological parameters from the Simulation~\#3 $(w_0=-0.8, w_a=0.1)$ and its phantom twin $(w_0=-1.2, w_a=-0.1)$ for $c_{\rm s}^2 = 10^{-7}$. Different columns present the comparison at the selected redshifts $z\in\{0.0, 0.3, 0.9\}$.}
    \label{fig:newmodel-phantom}
\end{figure*}

\begin{figure}
    \centering
    \includegraphics[width=0.99\columnwidth]{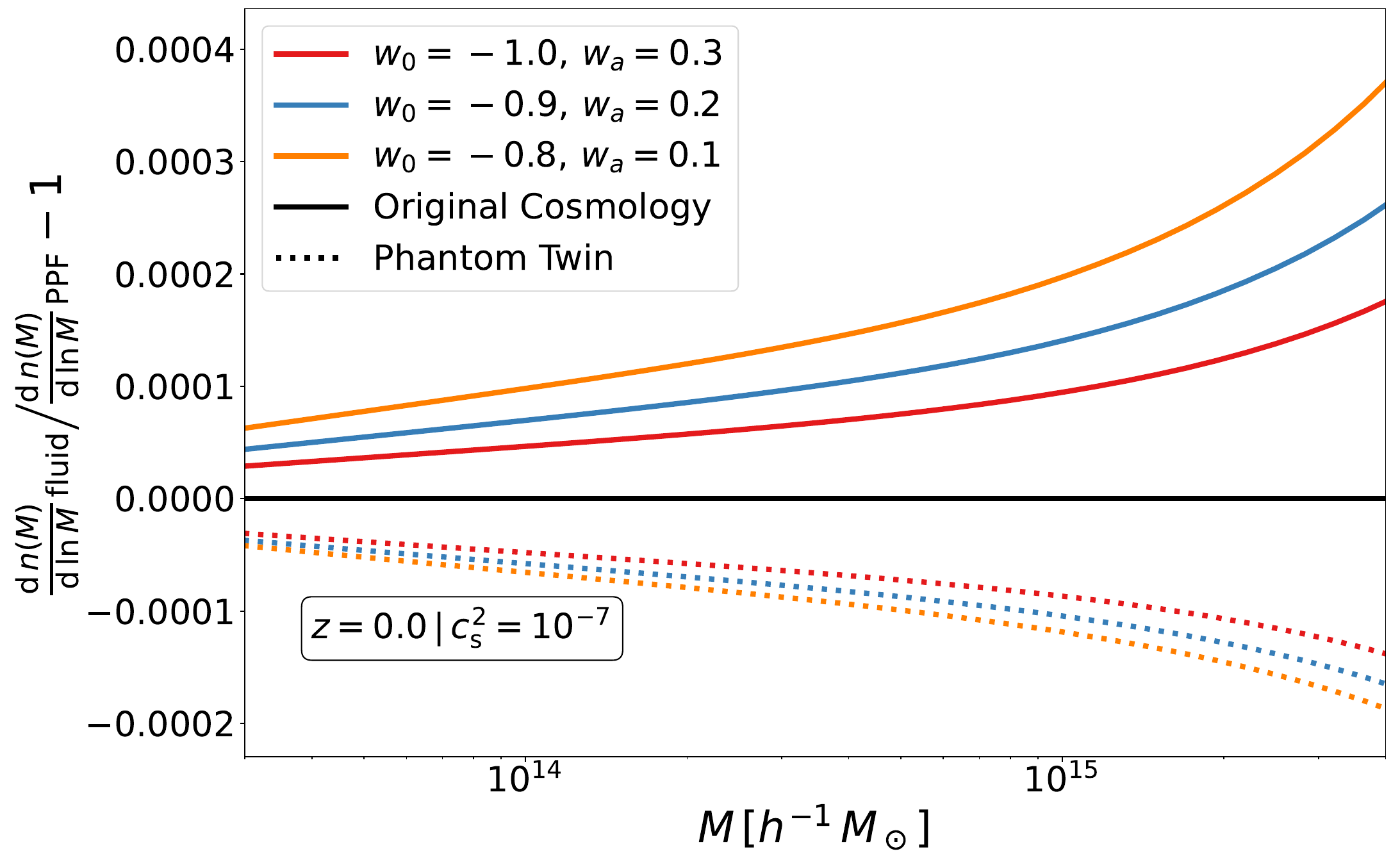}
    \caption{Relative difference in the HMF between the fluid and PPF approaches plotted as a function of the halo mass. Results are shown at redshift $z=0$ for the cosmologies corresponding to simulations~\#1--3. The solid lines denote the original cosmology, while dotted lines indicate the corresponding phantom twin.}
    \label{fig:fluidvsppf}
\end{figure}

In Fig.~\ref{fig:newmodel}, we present the model calibrated on the Simulations 1--12 using the PPF description presented in Table~\ref{tab:simulations}. The best fit for $\lambda$ obtained by maximizing the likelihood presented in Sect.~\ref{sec:cal} is $\lambda = 0.34$. We have not observed a statistically significant evolution of $\lambda$ with redshift or $c_{\rm s}$. We compare the model prediction against the simulations for $c_{\rm s}^2 \in \{10^{-7}, 10^{-5}\}$ as larger sound speeds resulted in an effect that is too small as we show in Fig.~\ref{fig:cs2}. Different rows present the comparison at various redshifts. 

From Fig.~\ref{fig:newmodel}, we observe that the relative impact of CDE is below one percent for all scenarios considered for redshift $z\gtrsim1$. From the plots of the residuals, we note that the calibrated model reproduces the simulation with sub-percent accuracy, except in the high mass end, where the count errors become evident. Since the simulations share the same seed, the ratio concerning the $c_{\rm s}^2=1$ case suppresses the Poisson noise, making the small impact of the CDE assessable. Still, when the number of halos inside a mass bin becomes small, the ratio of the simulations presented in Fig.~\ref{fig:newmodel} starts fluctuating more significantly around the model.

In Fig.~\ref{fig:newmodel-phantom}, we present our model prediction for the simulation~\#3 and its phantom twin. While the former has been used to calibrate our model, the latter is a real prediction of our model. We observe that our model's performance is consistent for the simulation~\#3 and its twin. We also observe that the amplitude of the overall impact is not simply reflected about zero, with the phantom impact being smaller by a factor of two. This can be understood as the contribution of DE perturbations to the Poisson equation is driven by
\begin{equation}
    \delta \rho_{\rm DE}^{\mathrm{Nb}} = \delta \rho_{\rm DE}^{\mathrm{S/N}} + 3 \mathcal{H}(1 + w_{\rm DE}) \frac{\theta_{\mathrm{total}}^{\mathrm{S/N}}}{k^2} \overline{\rho}_{\rm DE}\,,
\end{equation}
where the super-scripts refer to $N$-body, synchronous, and Newtonian gauges, respectively. Thus, the contribution of the velocity divergence field $\theta_{\rm total}^{S/N}$ flips its sign when phantom crossing. Starting from adiabatic initial conditions, where density fluctuations of all species couple to the same curvature fluctuations, the phantom density perturbations $\delta \rho_{\rm DE}$ and divergence of the velocity field initially have different signs, causing a slower evolution with respect to the non-phantom twin. Moreover, in phantom models, the background energy density decays faster with redshift than in non-phantom models, which further limits the impact of phantom DE pertubations.

Lastly, in Fig.~\ref{fig:fluidvsppf}, we compare the differences in the HMF between the PPF and fluid descriptions. We present the predictions of the best-fit model for $c_{\rm s}^2=10^{-7}$ at redshift $z=0$ where the differences are the largest. Still, the predicted relative difference is the order of $10^{-4}$ below both the accuracy and precision our simulations can probe and the overall effect of CDE.

\section{Discussion}
\label{sec:discussion}

An interesting result of our analysis is that previous models used to assess the impact of CDE on the HMF tended to predict a larger effect than what is observed in our simulations. For instance, \citet{Batista:2017lwf} predicted that CDE could affect the abundance of clusters by up to 30 percent. In contrast, our simulations reveal a maximum relative difference of only a few percent at the mass scale of $10^{15}\,h^{-1}\,M_\odot$.

To reconcile these divergent results, it is important to note that \citet{Batista:2017lwf} employed the spherical collapse model framework to describe the nonlinear evolution of DE, assuming a null sound speed. This approach presumes that the nonlinearities between DE and matter remain correlated throughout their co-evolution. However, as demonstrated in~\citet{Nouri-Zonoz:2024dph}, the correlation between the two fields decays rapidly once DE becomes nonlinear. Moreover, as shown in \citet{Batista:2022ixz}, a value of $c^2_{\rm s} \le 10^{-7}$ can be considered effectively as a null sound speed at cluster scales, leading to significant DE contrasts near halo virialization overdensities. Therefore, given that our current analysis assumes linear DE perturbations, our results should be regarded as a lower limit on the impact of DE fluctuations. Likely, the accurate effect for $c^2_{\rm s}\sim10^{-7}$ lies between our method and the predictions based on the spherical collapse approach.

Another interesting effect is that CDE is enhanced in non-phantom DE scenarios. In our analysis, models with an EOS that remains above the phantom divide exhibit a more pronounced impact on the halo mass function. This behavior is expected because DE perturbations contribute more effectively to the gravitational potential when $w_{\rm DE} > -1$.

It is also noteworthy that CDE could potentially induce an effect analogous to the tension observed in $\sigma_8$ between CMB and LSS surveys~\citep{Battye:2014qga,Douspis:2018xlj}. In fact, DE perturbations can either boost or deboost the matter fluctuations, making the collapse of structures easier or harder. The signal of the impact, as depicted in Eq.~\eqref{eq:sigma_mod} and Fig.~\ref{fig:densityfield}, depends on the DE EOS. 

However, the overall impact remains very modest for the scenarios simulated with our current numerical setup. For instance, in our simulation ~\#3, which has the largest CDE impact, at redshift $z=0$, the bias between the fiducial value of $A_{\rm s}$ and the one that reproduces the number of halos more massive than $10^{14}\,h^{-1}\,M_\odot$ ignoring CDE is below one percent. The bias reaches roughly one percent if we extrapolate our model to $c_{\rm s}^2=0$, but the behavior of our model beyond its validity is not easily assessable and, therefore, not encouraged. A different approach would be necessary to reliably explore the extreme case of $c_{\rm s}^2 = 0$.

Finally, the variations stemming from assuming the fluid or PPF description for DE are much smaller than the already limited impact of CDE. This finding suggests that uncertainties in DE microphysics are unlikely to introduce significant biases in the predictions for halo abundances when CDE's dominant contributions are adequately considered.

\section{Conclusions}
\label{sec:conclusions}

We have presented an improved HMF model that explicitly incorporates the impact of CDE with a variable sound speed. Our analysis is based on an extended suite of $N$-body simulations covering a wide range of DE parameters and sound speeds. The main conclusions of our study are as follows:

\begin{itemize}
    \item The effect of CDE on the HMF depends sensitively on the DE EOS. In models with $w_{\rm DE} > -1$, DE perturbations enhance gravitational collapse, whereas in phantom scenarios ($w_{\rm DE} < -1$) the opposite effect is observed.
    
    \item The impact of CDE is more pronounced in non-phantom DE scenarios, where DE perturbations contribute more effectively to the gravitational potential.
    
    \item When the matter power spectrum computed for $c_{\rm s}^2<1$ is applied directly to the baseline model of Paper~I, the qualitative behavior agrees with the simulation outcomes; however, the impact is overpredicted by roughly a factor of two.
    
    \item We argue that this overprediction arises because DE clusters much less efficiently than matter. Therefore, extrapolating the impact of CDE from the linear power spectrum exaggerates its effect on the non-linear collapse of structures.
    
    \item To include the effect of CDE on halo collapse, we introduce an effective peak height [Eq.~\eqref{eq:sigma_mod}] that is defined in terms of the matter power spectrum for the $c_{\rm s}^2=1$ case and modulated by the relative contribution of CDE perturbations at turnaround. This formulation more accurately captures the interplay between DE and matter during collapse.
    
    \item Although CDE could, in principle, generate an effect analogous to the $\sigma_8$ tension observed between CMB and large-scale structure measurements, the overall impact remains modest for the range of sound speeds explored in our simulations. Extrapolation to $c_{\rm s}^2=0$ suggests a larger bias; however, this extreme regime would require an alternative numerical treatment as the linear realization methodology will not be valid if CDE starts to deviate strongly from linearity.
    
    \item Finally, the difference between the PPF and fluid approaches for DE is much smaller than the limited impact of CDE itself. Hence, uncertainties in DE microphysics are unlikely to introduce significant biases in halo abundance predictions when the dominant effects of CDE are properly modeled.
\end{itemize}

Our extended HMF model maintains sub-percent accuracy in reproducing simulation outcomes and provides a robust framework for interpreting forthcoming galaxy cluster surveys. Future work should focus on alternative simulation approaches to reliably probe the extreme $c_{\rm s}^2=0$ regime and further refine the model in light of next-generation observational data.

\section{Data availability}
\label{sec:data}

The baseline model used in this paper is implemented in~\citet{Castro_CCToolkit_A_Python_2024}.\footnote{\url{https://github.com/TiagoBsCastro/CCToolkit}} The code to compute DE power-spectrum is implemented in~\citet{tiago_castro_2025_15187970}.\footnote{\url{https://github.com/TiagoBsCastro/ConceptSpectra}}

\begin{acknowledgements}

It is a pleasure to thank Duca dos Anjos. We warmly thank Francesco Pace for sharing solutions to the spherical collapse with us. We thank Isabella Baccarelli, Fabio Pitari, and Caterina Caravita for their support with the CINECA environment. We acknowledge the CINECA award for the availability of high-performance computing resources and support as part of the Leonardo Early Access Program (LEAP) and under the ISCRA initiative. RCB, TC, VM also thank Conselho Nacional de Desenvolvimento Científico e Tecnológico -- CNPq (process 444368/2024-8) for partially supporting this work. 
TC and SB are supported by the Agenzia Spaziale Italiana (ASI) under - Euclid-FASE D Attivita' scientifica per la missione - Accordo attuativo ASI-INAF n. 2018-23-HH.0, by the National Recovery and Resilience Plan (NRRP), Mission 4, Component 2, Investment 1.1, Call for tender No. 1409 published on 14.9.2022 by the Italian Ministry of University and Research (MUR), funded by the European Union – NextGenerationEU– Project Title "Space-based cosmology with Euclid: the role of High-Performance Computing" – CUP J53D23019100001 - Grant Assignment Decree No. 962 adopted on 30/06/2023 by the Italian Ministry of University and Research (MUR); by the Italian Research Center on High-Performance Computing Big Data and Quantum Computing (ICSC), a project funded by European Union - NextGenerationEU - and National Recovery and Resilience Plan (NRRP) - Mission 4 Component 2, by the INFN INDARK PD51 grant, and by the PRIN 2022 project EMC2 - Euclid Mission Cluster Cosmology: unlock the full cosmological utility of the Euclid photometric cluster catalog (code no. J53D23001620006). VM acknowledges partial financial support FAPES (Brazil).

\end{acknowledgements}

\bibliography{mybib}

\end{document}